# First Production of New Thin 3D Sensors for HL-LHC at FBK

**D M S Sultan**[a,b,*], **Gian-Franco Dalla Betta**[a,b], **Roberto Mendicino**[a,b], **Maurizio Boscardin**[c,b], **Sabina Ronchin**[c,b], **and Nicola Zorzi**[c,b]

[a] *DII, University of Trento, Via Sommarive, 9, 38123 Trento, Italy*
[b] *TIFPA – INFN, Via Sommarive, 14, 38123 Trento, Italy*
[c] *Fondazione Bruno Kessler (FBK-CMM), Via Sommarive, 18, 38123 Trento, Italy*

E-mail: *dms.sultan@unitn.it*

ABSTRACT: Owing to their intrinsic (geometry dependent) radiation hardness, 3D pixel sensors are promising candidates for the innermost tracking layers of the forthcoming experiment upgrades at the "Phase 2" High-Luminosity LHC (HL-LHC). To this purpose, extreme radiation hardness up to the expected maximum fluence of $2\times10^{16}$ $n_{eq}.cm^{-2}$ must come along with several technological improvements in a new generation of 3D pixels, i.e., increased pixel granularity (50×50 or 25×100 μm² cell size), thinner active region (~100 μm), narrower columnar electrodes (~5μm diameter) with reduced inter-electrode spacing (~30 μm), and very slim edges (~100 μm). The fabrication of the first batch of these new 3D sensors was recently completed at FBK on Si-Si direct wafer bonded 6" substrates. Initial electrical test results, performed at wafer level on sensors and test structures, highlighted very promising performance, in good agreement with TCAD simulations: low leakage current (<1 pA/column), intrinsic breakdown voltage of more than 150 V, capacitance of about 50 fF/column, thus assessing the validity of the design approach. A large variety of pixel sensors compatible with both existing (e.g., ATLAS FEI4 and CMS PSI46) and future (e.g., RD53) read-out chips were fabricated, that were also electrically tested on wafer using a temporary metal layer patterned as strips shorting rows of pixels together. This allowed a statistically significant distribution of the relevant electrical quantities to be obtained, thus gaining insight into the impact of process-induced defects. A few 3D strip test structures were irradiated with X-rays, showing inter-strip resistance of at least several GΩ even after 50 Mrad(Si) dose, thus proving the p-spray robustness. We present the most important design and technological aspects, and results obtained from the initial investigations.

---

[*] D M S Sultan (*Corresponding Author*)

# Contents

## 1. Introduction

The upgrades of the major experiments (ATLAS and CMS) at the High-Luminosity LHC (HL-LHC) aim at the complete replacement of their tracking detectors to cope with the predicted higher event pile-up (140 events/bunch-crossing) and integrated luminosity of 2500 fb$^{-1}$. The innermost tracking layers will have to withstand very large radiation fluences up to $2 \times 10^{16}$ 1-MeV equivalent neutrons per square centimeter ($n_{eq}$ cm$^{-2}$ ) [1]. Owing to their intrinsic (geometry dependent) radiation tolerance [2], 3D sensors are very promising candidates for this application. However, despite their remarkable performance, existing 3D pixel sensors, e.g., those currently installed in the ATLAS IBL [3,4], cannot fulfill the challenging demands of HL-LHC, calling for the development of a new generation of these devices which should feature: very dense pixel granularity (50×50 or 25×100 µm$^2$), thinner active region (~100 µm), narrower columnar electrodes (~5µm diameter) with reduced inter-electrode spacing (~30 µm), and very slim edges (~100 µm).

Fondazione Bruno Kessler (FBK) of Trento, Italy, was one of the two fabrication facilities, together with CNM-IMB of Barcelona, Spain, involved in the 3D pixel sensor production for the ATLAS IBL [4]. Since then, the fabrication line at FBK was upgraded to process 6-inch diameter silicon wafers [5]. While the increased area available on wafer (about a factor of two as compared to 4-inch diameter) can offer a significant advantage in case of volume productions, maintaining the double-sided fabrication process [6] for thin 6-inch substrates is not the best choice because of mechanical issues (compatibility with processing equipment, fragility). Therefore, a different technological approach has been conceived, that uses Si-Si direct wafer bonded (DWB) substrates with a single-sided process [7]. The fabrication of a first batch of these new, small-pitch 3D sensors was recently completed at FBK, and extensive electrical measurements were carried out with encouraging results.

In this paper, we report on the fabrication technology and the design of these new 3D sensors, as well as on selected results from the electrical characterization of different sensors (diodes, strips, pixels), also in comparison to TCAD simulation predictions. Preliminary results from the electrical characterization of strip sensors irradiated with X-rays up to 50 Mrad(Si) dose are also reported.

## 2. Fabrication of modified single-sided 3D sensors

The devices presented in this paper are the first small-pitch 3D sensors with thin active layer ever fabricated. To this purpose, we used a single-sided process on Si-Si DWB substrates from IceMOS Technology Ltd. These substrates consist of a Float Zone high-resistivity layer of the desired thickness (namely 100 µm and 130 µm in this first production) directly bonded (i.e., without an oxide layer in between) to a 400-µm thick low-resistivity handle wafer, thus ensuring sufficient mechanical robustness. The quality of this kind of raw material, which was previously unexplored for detector applications, was assessed by a preliminary batch of planar sensors [8]: from leakage current measurements, the bulk generation lifetimes were proved to be quite good, of the order of a few ms. Moreover, from capacitance-voltage curves of test diodes, it was possible to extract the substrate doping profiles, which revealed a significant back-diffusion of boron dopant from the handle wafer into the active layer with a depth of about 10 μm. This makes the effective thickness of the active layer 10 μm smaller, a fact that should be kept in account for the etching of the columnar electrodes.

## 2.1 Process design

The schematic cross-section of the proposed sensors is shown in Fig.1. The direct bonding of the high Ω-Si (active) layer (<100> crystal orientation, resistivity higher than 3kΩ cm) to the low Ω-Si handle wafer dictates the choice of different column depths for the junction ($n^+$) and the ohmic ($p^+$) electrodes. The p-columns are etched first by Deep Reactive Ion Etching (DRIE) through the whole high Ω-Si layer thickness and penetrate by a few μm into the low Ω-Si layer. By doing so, a good ohmic contact is made on the handle wafer, that can eventually be partially removed with a post processing and coated with a metal layer to allow for sensor bias from the back-side. Lessons learned from the ATLAS IBL experience suggest this is essential to ensure 3D sensor installation within a pixel module [3,4]. On the contrary, the n-columns do not penetrate through the entire active layer, but rather stop at a short distance (~25 μm) away from the low Ω-Si layer, in order to avoid early breakdown. A p-spray layer obtained with a low dose Boron implantation is present at the front side, to prevent the inversion of the surface, thus isolating the n-columns [5].

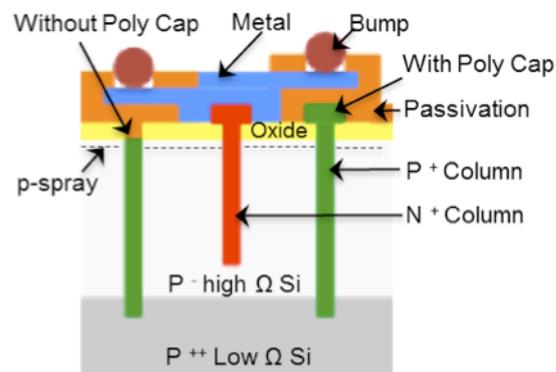

**Fig. 1**: Schematic cross-section of the proposed single-sided 3D sensors on Si-Si DWB substrates.

Both types of columns are partially filled with doped poly-Si, that is deposited also at the surface. As shown in Fig. 1, p-columns come with two possible variants relevant to the removal of the poly-Si from the front side surface, i.e., "with poly-cap" (right side in Fig. 1), that is obtained by etching after defining the poly-Si patterns by lithography, and "without poly-cap" (left side in Fig.1), that is obtained by etching poly-Si everywhere without a mask. Table 1 summarizes the process splits used in the first batch. Among the 10 processed wafers, note that W82 was broken at the end of the process before temporary metal deposition.

**Table 1**: Processed wafer summary with different options.

|  | Wafer ID | | | | | | | | | |
|---|---|---|---|---|---|---|---|---|---|---|
|  | W36 | W41 | W48 | W50 | W54 | W70 | W76 | W77 | W78 | W82 |
| **Poly-cap presence** | yes | | | no | | | yes | | no | |
| **Nominal Active Wafer thickness [μm]** | 100 μm | | | | | | 130 μm | | | |

## 2.2 Process characterization

The process described above come with the reduction of the processing steps (almost 30% less than the IBL technology) but also faces the challenges relevant to the single-sided process of both types of columns with a small pitch. Figure 2a shows the layouts of the two most important pixels cells, featuring 50×50 μm$^2$ and 25×100 μm$^2$ sizes. The former has one n$^+$ column (1E) at the center, with an inter-electrode spacing L~36 μm. This layout is pretty safe since the distances between the different structures are large. In particular, the bump-bonding pad can be placed on either side of the n-column while remaining far enough (>15 μm) from the p-columns. On the contrary, the 25×100 μm$^2$ layout can either be made with one n-column (1E), with an inter-electrode spacing L~51 μm, or, for higher radiation hardness, with two n-columns (2E), with an inter-electrode spacing L~28 μm. In the latter case (Fig. 2a), the layout is really dense and the bump-bonding pad is very near to both the n- and the p-columns, so that misalignment problems could cause the pad to overlap the columns, thus leading to high electric fields with the risk of microdischarges. In this respect, the p-column finishing option "without poly-cap" would add a safety margin of a few μm, but it should be proved not to worsen the sensor electrical characteristics.

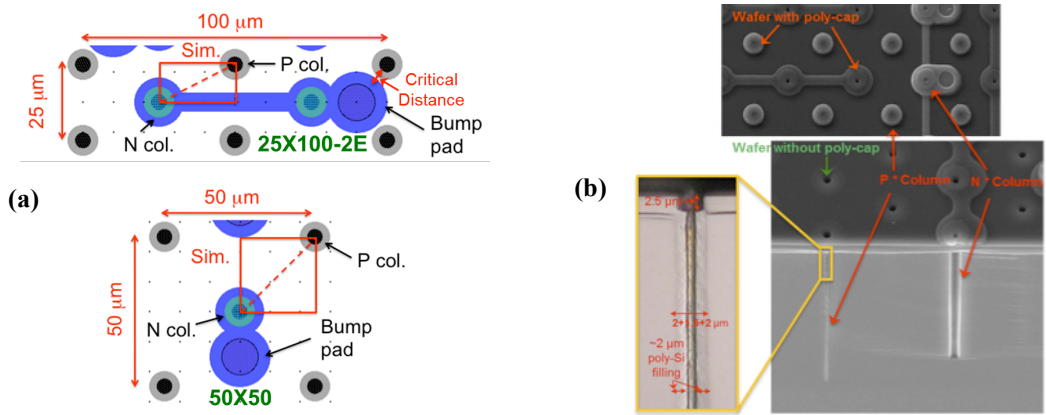

**Fig. 2**: **(a)** Layout-view of the 25x100-2E (up) and 50x50 (down) pixels; **(b)** SEM images: n$^+$ column and p$^+$ column with representation of both "with poly-cap" and "without poly-cap" options.

Like for every 3D sensor fabrication, the most delicate steps within this production were the column etchings by DRIE. The etching recipes had been previously optimized with dedicated tests [6], so that results were good. Figure 2b shows a few pictures from Scanning Electron Microscope (SEM). It can be seen that the column widths are pretty uniform along their depths. The micrographs also show that the poly-Si etching step "without cap" seems not to have damaged the p-columns.

## 3. Wafer Layout

The large area available on 6-inch wafers allows accommodating a huge variety of sensors on the wafer layout (called 3DSS-6, see Fig. 3a) while ensuring a wide safety margin (>1 cm) from the wafer edge. The core of the wafer layout contains pixel sensors compatible with the ATLAS FE-I4 and the CMS PSI46 read-out chips (ROCs). There are 13 "FE-I4" single tiles: some of them have the same column configuration as the IBL pixels [3], with or without guard ring, whereas all the others have new small-pitch pixels with 50×50, 25×100-2E, and 25×100-1E, with or without guard ring. There are 38 "PSI46" single tiles, differing in the number of n-columns (2 and 3) per pixel of conventional geometries, along with the new small-pitch designs, with or without guard ring. There are also 9 "FE-I3" complaint sensors with a combination of new pixel sizes. Some other sensors are instead compatible with the new ROCs being developed by the RD53 Collaboration [9]. There are 11 big sensors (0.9 cm$^2$) and 18 small sensors (0.36 cm$^2$). All the big sensors have pixel cell size of 50×50 μm$^2$, whereas the small ones have either 50×50, or 25×100-2E, or 25×100-1E μm$^2$ cell size geometries.

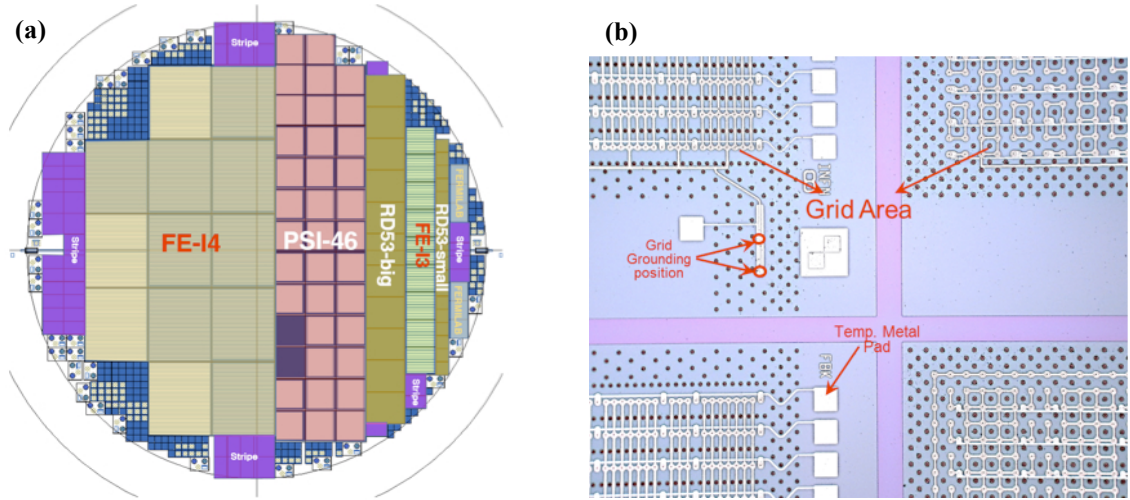

**Fig. 3**: **(a)** Wafer layout of the 3DSS-6 batch, with an indication of the many different sensors **(b)** Micrograph of pixel sensors, showing the metal grid shorting non-read-out pixels and the temporary metal.

All the pixelated sensors are designed with slim edges [10]. Compared to the 200-μm design used for the ATLAS IBL pixels, the possibility to use a reduced pitch for the columns here allows for more aggressive designs, down to 75-μm size [11]. The periphery of the wafer layout hosts several smaller sensors (3D strips, 3D diodes) and test structures. In particular, 3D diodes (~2 mm$^2$ area) reproduce the basic 3D cell layouts of all types of large pixel and strip sensors, but all columns of the same type are shorted together by a metal grid to obtain a 2-electrode device [6].

One of the biggest challenges of the layout design for this batch was to make the new, small-pitch pixels compatible to the existing ROCs that have larger pixel pitches. To this purpose, in the sensor layouts, n- and p-columns are placed on either 25×100 μm$^2$ or 50×50 μm$^2$ grids, corresponding to the elementary cells. One or more cells are then connected to the ROC bonding pads, whereas the remaining n-columns are all shorted by a metal grid and connected to the extra bonding pads that are grounded in the ROC (see Fig. 3b). This solution allows as many small pixels as possible to be tested, avoiding problems in ROC tuning, since all read-out channels have the same input capacitance, and ensuring proper boundary conditions, since all columns are uniformly biased. However, it should be noted that the geometrical efficiency is small (as an example, it is 20% in an FE-I4 based design, where the ROC pixel cells are 50×250 μm$^2$), making data analysis more difficult. Finally, Fig. 3b also shows the temporary metal grid that is used for on-wafer electrical tests of pixel sensors [6].

## 4. Simulation and electrical characterization

In order to predict and optimize the performance of these new 3D sensors, numerical device simulations have been performed using Synopsis TCAD software. Because of symmetry considerations, minimum size cells were considered as simulation domains (see red regions in Fig. 2a), while values of all parameters are representative of FBK technology. Full 3D simulations were performed, also incorporating detailed information about the dielectric layers, the metal layers, and the surface insulating layer (p-spray). The column diameters are 5 μm, the active layer thickness is 150 μm and the n-column depth is 130 μm. From I-V simulations (Fig. 4a), the leakage current is small and, most notably, the breakdown voltage is large (~150 V). Inspection of electric field distribution at a voltage close to breakdown yields similar peak values at the surface junction and at the n-column tips. From the knees in C-V curves (Fig. 4b), a very small depletion voltage (a few V) can be estimated; moreover, the saturation value of the capacitances is ~50 fF per column, making the total value for the 25x100-2E pixel close to the RD53 recommended limit of 100 fF [9].

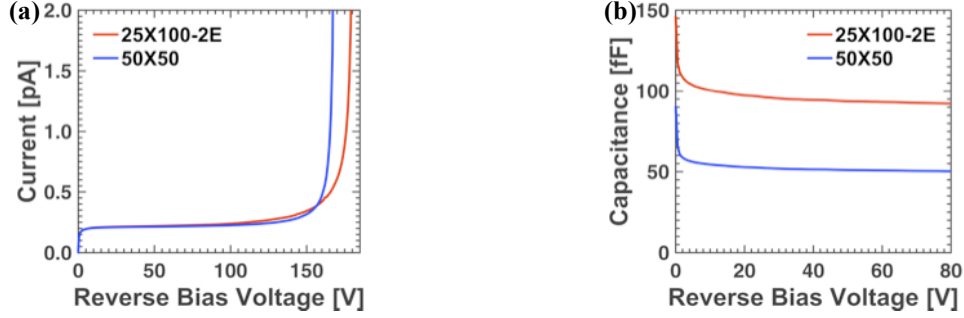

**Fig. 4**: Simulated electrical characteristics for 50x50 and 25x100-2E pixels **(a)** I-V curves, **(b)** C-V curves.

Many test structures, mainly 3D diodes, were probed with the experimental setup for electrical characterization at room temperature, using the required combination of HP4284A LCR meter at the frequency of 10 kHz, HP4145B semiconductor parameter analyzer with average integration time and Keithley 2410 voltage source. For all considered 3D cell dimensions, test diodes are available with or without a guard ring; as an example, Fig. 5a shows the layout of a 3D diode with guard ring. As can be seen from Fig. 5b, the leakage currents are small: the measured values, if normalized to the number of columns, are in the range from 0.3 to 3 pA. The relatively wide spread of the current values is likely to be ascribed to the different quality of the active layer material for 100-µm and 130-µm thick wafers in terms of carrier lifetimes, as also observed in a preliminary test batch of planar devices [8]. In Fig. 5b it can also be observed that the breakdown voltage is large (~150 V), comparable to simulated values. The C-V curves of 3D diodes of different thickness and featuring the two most important cell layouts are reported in Fig. 5c, that confirm the very low depletion voltage of these sensors. After normalizing the capacitance values to the respective number of columns present in the diodes, results are found to be of the order of 50 fF per column, very close to the simulated values, and with slight differences (4-5 fF per column) between the two different active layer thicknesses.

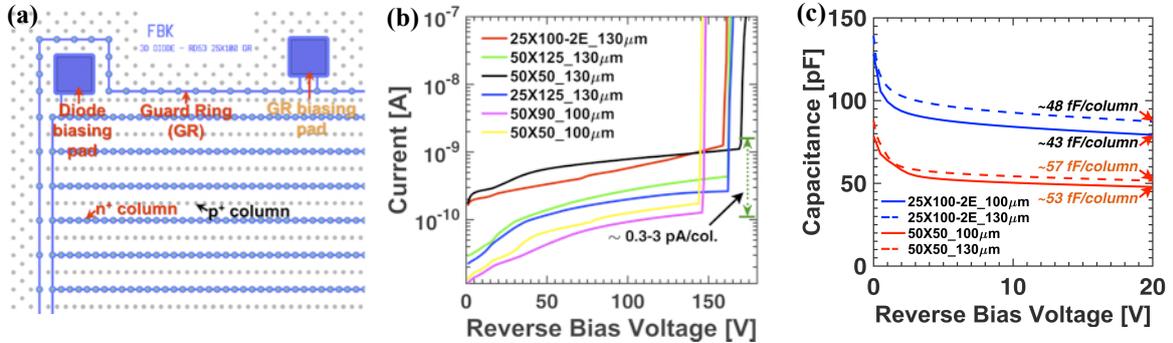

**Fig. 5**: **(a)** Layout of a 3D diode with guard ring; **(b)** I-V curves of diodes of different active thickness and unit cell geometries; **(c)** C-V curves of diodes of different active thickness and unit cell geometries.

The I-V curves of all pixel sensors were initially measured on wafer with an automatic probe station making use of a temporary metal layer patterned in strips that short-circuit rows of pixels together [6]. By doing so, it is possible to monitor the quality of pixel sensors before bump bonding by means of statistically significant distributions of the main parameters. As an example, Fig. 6 shows results relevant to the so called "RD53-big" sensors, which feature an array of 200x168 pixels of 50×50 µm$^2$ size. Sensor from wafers with both 100-µm and 130-µm active layer thickness are considered, also evaluating the impact of the poly-cap option.

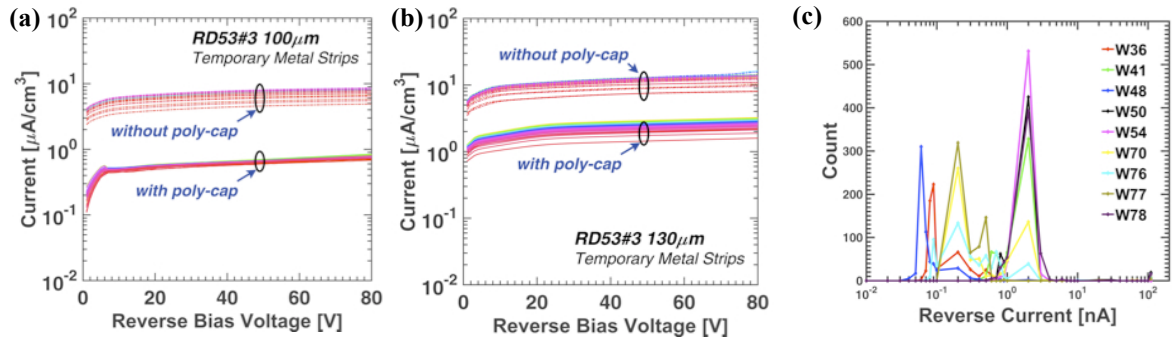

**Fig. 6**: Leakage current measurements through temporary metal strips, each one short-circuiting 3 rows of 200 pixels (50x50 μm²) in "RD53-big" pixel sensors: **(a)** I-V curves of a few strips with 100-μm active thickness with poly-cap (W36) and without poly-cap (W54); **(b)** I-V curves of a few strips with 130-μm active thickness with poly-cap (W76) and without poly-cap (W78); **(c)** Distribution of the currents at full depletion (20 V) of all strips from all sensors (11 per wafer) on 9 wafers.

Figures 6a and 6b show I-V curves for selected samples of metal strips belonging to sensors from different wafers. Note that every strip shorts 3 rows of 200 pixels each. Similarly to diode measurements, it is confirmed that leakage currents are lower for substrates with 100-μm thick active layer. The observed difference in the currents is in fact much larger (~3.5x) than expected from the different thickness alone. Most notably, for both 100-μm and 130-μm thick active layers, the option "without poly-cap" exhibits much larger leakage currents (~10x). This difference seems to indicate that the removal of the poly-Si layer from the wafer front-side without a mask actually induces some damage at the p-column openings, although this was not evident from SEM micrographs. The distribution of leakage current values measured at 20 V reverse bias (full depletion) in all RD53-big sensors (11 per wafer, from 9 wafers) is reported in Fig. 6c. The wide dispersion of leakage current values is evident also from this plot, however it should be noted that even the largest values, if normalized to the number of columns, correspond to just a few pA, that is remarkably low for 3D sensor technology, and comparable to the best results obtained from the ATLAS IBL production [6]. From Fig. 6c, it can also be observed that the total number of strips with defects causing high current (indicated by the small peak at ~100 nA in the distribution), is 81/5544, corresponding to ~1.46%. This small value demonstrates the good process quality even for relatively big devices with high column density, and is promising in view of future productions.

In 3D sensors from previous productions at FBK, carried out with the double-sided process approach, the breakdown voltage values were found to be largely dependent on the properties of the Si/SiO$_2$ interface, and particularly the oxide charge and interface state densities [12]. In order to assess this aspect for the new technology, two sets of strip sensors including three different small-pitch designs, i.e., 50x50, 25x100-1E, and 25x100-2E, and one large pitch design (80x80), were irradiated with 10-keV X-rays at the Rutherford Appleton Laboratory (UK) to two Total Ionizing Doses (TIDs): 5 Mrad(Si) and 50 Mrad(Si). During irradiation, devices were not biased. Electrical tests were performed on a significant fraction of strips in each sensor.

Fig. 7a shows the I-V curves of one strip in a 50x50 sensor from W82 (130-μm active thickness, without poly-cap) in different conditions: before irradiation, after 5 Mrad(Si), and after 50 Mrad(Si). It can be seen that X-ray irradiation caused a sizable increase of the leakage current, due to surface generation, but only minor changes in the breakdown voltage. Fig. 7b shows the I-V curves for different designs at different TIDs. Also in this case, negligible differences in the breakdown voltage are observed. However, it is possible to appreciate the much high impact of X-ray irradiation on the leakage current in case of designs with higher densities of columns.

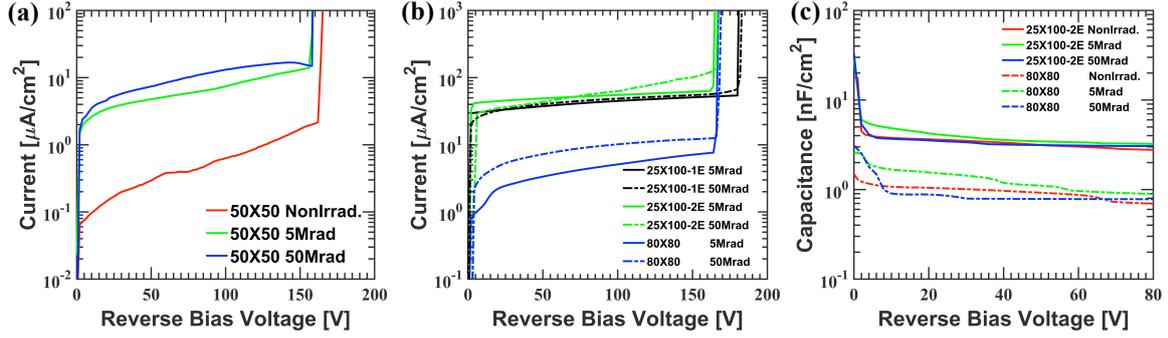

**Fig. 7**: Electrical characteristics of strip sensors of 130-μm active thickness without poly-cap (W82): **(a)** I-V curves of 50×50 μm² pixel size before and after irradiation at different TIDs **(b)** I-V curves of different pixel geometries at different TIDs, and **(c)** C-V curves of different pixel geometries before and after irradiation at different TIDs.

The leakage current dependence on the column density can be explained by the surface generation contribution that originates from the depleted regions at the n-column/p-spray junctions on the front side: the higher the number of columns, the larger the effective depleted perimeter, the larger the current. On the contrary, the weak dependence of the breakdown voltage on the irradiation is less obvious. Although direct measurements could not be performed, X-ray irradiation is expected to have caused a significant increase of oxide charge and interface state densities as compared to the pre-irradiation values. As a result, the breakdown voltage value should be increased if breakdown occurs at the surface on the front side [12]. Conversely, the minor changes observed in the breakdown voltage after irradiation suggest this is not the case, and that a different mechanism, independent of surface damage, limits the breakdown voltage, most likely due to the high electric field at the n-column tips.

Figure 7c shows the C-V curves for different designs at different TIDs. As expected, the capacitance is much higher for the design with higher column density. Moreover, X-ray irradiation seem to have a minor impact on the capacitance, that is compatible with a prevalent contribution from the columnar electrodes rather than from the surface.

Finally, interstrip resistance measurements were performed (not shown). The important outcome of these tests is that the interstrip resistance remains very large (at least a few GΩ) even for the largest TID regardless of the specific design. This confirms the effectiveness of the p-spray layer in isolating the n-columns also in the presence of severe surface damage.

## 5. Conclusion

We have reported on the first small-pitch, thin 3D sensors produced at FBK and oriented to the HL-LHC upgrades. Devices were fabricated on 6-inch Si-Si direct wafer bonded substrates with a single-sided process, for which the most critical steps of column etching and filling with poly-Si were previously tested. In spite of the very high column densities, these new 3D sensors were proved to be properly functioning with very good electrical characteristics: among them, very small leakage currents (~pA per column), high breakdown voltages (~150 V), and small capacitances (~50 fF per column). All these values are in good agreement with TCAD simulation, thus assessing the quality of the design approach and of the technology.

Extensive measurements of the I-V curves of pixel sensors by means of a temporary metal layer allowed to obtain statistically significant distributions of the leakage current values, from which a remarkably low defect density was found. Taking into account this is the first prototype run made with the new single-sided technology, this is very encouraging in view of future productions.

X-ray irradiation of strip sensors up to a TID of 50 Mrad(Si) confirmed the effectiveness of the p-spray layer in isolating the n-columns at the surface, also in the presence of large oxide charge and interface state densities. The negligible variation of the breakdown voltage with TID also pointed to the high electric field developing at the n-column tips as probable origin of breakdown in these devices. However, this is not a major concern since the relatively large breakdown voltage here reported are believed to be high enough for efficient operation of these small-pitch sensors even after large irradiation fluences.

## Acknowledgment

The authors would like to thank Dr. John Matheson, Rutherford Appleton Laboratory (UK), for his support with X-ray irradiation of 3D strip sensors. Fruitful discussions with Dr. Marco Meschini (INFN Firenze), Dr. Giovanni Darbo (INFN Genova), and Prof. Alberto Messineo (INFN and University of Pisa) are also warmly acknowledged.

This project has received funding from the European Union's Horizon 2020 Research and Innovation programme under Grant Agreement no. 654168. This work was also supported by the Autonomous Province of Trento through the Project MEMS3, and by the Italian National Institute for Nuclear Physics (INFN) through Projects ATLAS, CMS, and RD-FASE2 (CSN1).